# On the effect of heterovalent substitutions in ruthenocuprates

P.W. Klamut,[a,b,f,*] B. Dabrowski,[a,b] S. M. Mini,[a,b] M. Maxwell,[a] J. Mais,[a]
I. Felner,[c] U. Asaf,[c] F. Ritter,[c]
A. Shengelaya,[d] R. Khasanov,[d] I. M. Savic,[d] H. Keller,[d]
A. Wisniewski,[e] R. Puzniak,[e] I. M. Fita,[e]
C. Sulkowski,[f] M. Matusiak[f]

[a] *Department of Physics, Northern Illinois University, DeKalb, Illinois 60115, USA*

[b] *Materials Science Division, Argonne National Laboratory, Argonne, USA*

[c] *Racah Institute of Physics, The Hebrew University, Jerusalem, Israel*

[d] *Physik-Institut der Universität Zürich, CH-8057 Zürich, and Paul Scherrer Institute, Villigen, Switzerland*

[e] *Institute of Physics of the Polish Academy of Sciences, 02-668 Warszawa, Poland*

[f] *Institute of Low Temperature and Structure Research of the Polish Academy of Sciences, 50-950 Wroclaw, Poland*

**Abstract**

We discuss the properties of superconducting derivatives of the $RuSr_2GdCu_2O_8$ (1212-type) ruthenocuprate, for which heterovalent doping has been achieved through partial substitution of Cu ions into the $RuO_2$ planes ($Ru_{1-x}Sr_2GdCu_{2+x}O_{8-\delta}$, $0 \leq x \leq 0.75$, $T_c^{max}=72$ K for $x=0.3-0.4$) and Ce ions into the Gd sites ($RuSr_2Gd_{1-y}Ce_yCu_2O_8$, $0 \leq y \leq 0.1$). The measurements of XANES, thermopower, and magnetization under external pressure reveal an underdoped character of all compounds. Muon spin rotation experiments indicate the presence of magnetic order at low temperatures ($T_m=14$-2 K for $x=0.1$-0.4). Properties of these two series lead us to the qualitative phase diagram for differently doped 1212-type ruthenocuprates. The difference in temperature of magnetic ordering found for superconducting and non-superconducting $RuSr_2GdCu_2O_8$ is discussed in the context of the properties of substituted compounds. The high pressure oxygen conditions required for synthesis of $Ru_{1-x}Sr_2RECu_{2+x}O_{8-\delta}$, have been extended to synthesis of a $Ru_{1-x}Sr_2Eu_{2-y}Ce_yCu_{2+x}O_{10-\delta}$ series. The Cu→Ru doping achieved in these phases is found to decrease the temperature for magnetic ordering as well the volume fraction of the magnetic phase.

[*] Corresponding author. Tel.: +1-815-753-6481; Fax: +1-815-753-8565; e-mail: klamut@niu.edu

## 1. Introduction

The properties of superconducting $RuSr_2GdCu_2O_8$ (1212-type) [1] and $RuSr_2RE_{2-y}Ce_yCu_2O_{10-\delta}$, RE=Gd, Eu (1222-type) [2,3] ruthenocuprates attracted a lot of interest after the possibility of microscopic coexistence of superconductivity (SC) and ferromagnetism (FM) was reported in these compounds [3-5]. The general problem of accommodating SC and FM in the same volume of the material echoes the original considerations of Ginzburg [6] and Matthias [7] of a mutually exclusive singlet state superconductivity and ferromagnetism. Activity remains revealing the different microscopic nature of coexistence for different classes of compounds. Microscopic coexistence was first observed in several low temperature f-electron superconductors [8]; for example, $ErRh_4B_4$, where the ferromagnetism observed between 0.9 and 1.4 K is modified to a spiral like structure to accommodate for the presence of superconductivity [9]. The superconducting phase also appears to be spatially inhomogeneous being interspersed with ferromagnetic domains. On the other hand, the recent discovery of superconductivity within a weak ferromagnetic state of an itinerant electron system in $UGe_2$ [10] suggests its association with ferromagnetic spin fluctuations and the spin triplet channel for pairing.

$RuSr_2GdCu_2O_8$ presents an interesting example of a layered perovskite-related structure, for which the anisotropy and strength of the expected interactions could in principle allow the accommodation of the superconducting and ferromagnetic order parameters [11]. Muon spin rotation spectroscopy experiments indicated ferromagnetism persisting in the superconducting state of this compound [5]. The results of neutron diffraction [12, 13] reported soon after, provided evidence of the antiferromagnetic (AFM) type coupling present in the Ru sublattice, and also set an upper limit of $0.1\mu_B$ to any ferromagnetic component. It should be noted, however, that an extra magnetic moment of approx. 1 $\mu_B$ is observed for the Ru sublattice at T=4.5 K and $\mu_oH$=7 T, i.e., well below $H_{c2}$ [14]. The hysteresis loop and irreversibility of magnetization vs. temperature observed below $T_N \cong 132$ K indicate weak-ferromagnetism that can originate in the canting of the Ru moments caused by the antisymmetric super-exchange interactions in the distorted structural block of the $RuO_6$ octhaedra. The latter resembles the behavior of the Cu spin system in the $Gd_2CuO_4$ weak-ferromagnet. In Ref. 15, the annealing induced modifications in the network of microstructural domains observed in $RuSr_2GdCu_2O_8$ have been associated with modification of the superconducting transition temperature. Appearance of fine structure of the superconducting phase has been suggested in Ref. 16. Reference 17 brings discussion of the possible effects of material's inhomogeneity based on recently reported Cu→Ru substituted phases [14]. Characterization of the microstructural details of the investigated samples seems to be of primary importance to proper understanding the peculiar properties of this material.

## 2. Results

Recently, we have reported the high-pressure oxygen synthesized (600bar at 1080°C) series of $Ru_{1-x}Sr_2GdCu_{2+x}O_{8-\delta}$ compounds where the Cu ions are partially doped into the $RuO_2$ planes of the 1212-type structure [14]. Increase of the Cu content in these materials was found to increase the hole doping (see Fig. 1 for the characteristic values of thermopower and the XANES Cu *K*-edge energy) and consequently the temperature of the superconducting transition up to $T_c=72$ K for x=0.4 [14, 18].

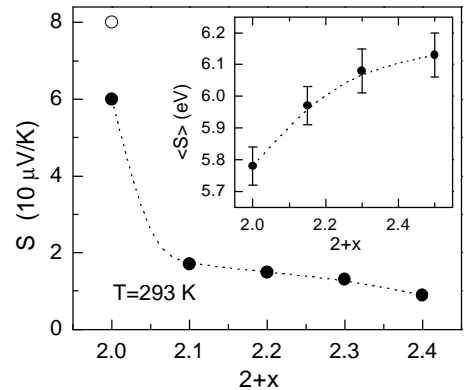

Fig. 1. Thermopower for $Ru_{1-x}Sr_2GdCu_{2+x}O_{8-\delta}$ superconductors as a function of Cu content. Open circle represents the value for non-superconducting $RuSr_2GdCu_2O_8$. Inset shows XANES Cu-K edge energy vs. Cu content, T=293 K.

On the other hand, previously reported [19], $Ce^{4+} \rightarrow Gd^{3+}$ substitution led us to the series of $RuSr_2Gd_{1-y}Ce_yCu_2O_8$ (0<y<0.1) where $T_c$ sharply decreases with y. By combining the properties of these two series of compounds we can construct a qualitative phase diagram for the family of 1212-type ruthenocuprates – see Fig. 2 (for detailed discussion see also Ref. 18). Heterovalent substitutions of $Nb^{5+}$ and $Sn^{4+}$ into the Ru site in $RuSr_2GdCu_2O_8$ also showed similar trend of change of $T_c$ with charge doping [20, 21].

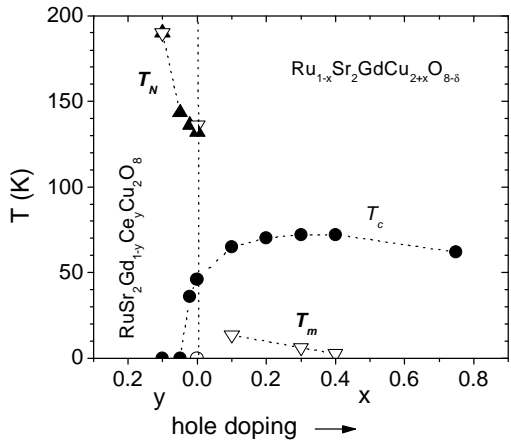

Fig. 2. Characteristic temperatures of $RuSr_2Gd_{1-y}Ce_yCu_2O_8$ and $Ru_{1-x}Sr_2GdCu_{2+x}O_{8-\delta}$ versus Ce→Gd and Cu→Ru substitutions. Open triangles: temperatures of magnetic phase transitions ($T_N$, $T_m$), as determined from temperature dependencies of the internal field measured in zero-field μSR experiment. Closed triangles and circles: temperatures of the magnetic ($T_N$) and superconducting ($T_C$) phase transitions [14, 18].

Interestingly, for x = 0.1 - 0.4 compositions of $Ru_{1-x}Sr_2GdCu_{2+x}O_{8-\delta}$ magnetic order at low temperatures has been recently observed in zero-field μSR experiments [22]. The origin of this effect needs to be investigated further. Its bulk character revealed in μSR measurements suggests inhomogeneity of the superconducting phase or microscopic coexistence of different phases at low temperatures. Alternatively, the magnetic stripes scenario supported by the results of μSR spectroscopy for several other underdoped

HTSC superconductors could be considered [23]. We note, however, that the magnetization characteristics reported for $Ru_{1-x}Sr_2GdCu_{2+x}O_{8-\delta}$ indicate the quasi-two dimensional character of the superconducting phase [14, 18]. For the x=0.4 sample, post synthesis annealing at 800°C performed in thermobalance revealed that, in resemblance to conventional 123-type superconductors, the $T_c$ can be varied by changing the oxygen content δ. Annealing in 0.01%, 21% and 100% of partial pressure of oxygen, led to the onset $T_c$=0, 47 and 52 K, and to the corresponding differences in the oxygen content Δδ=0.26, 0.32 (between samples with $T_c$=0 and $T_c$=47, 52 K).

Whereas for x≠0 the annealing induced change of $T_c$ can be explained by changes of δ, for x=0 both superconducting and non-superconducting $RuSr_2GdCu_2O_8$ have been reported, both in the form of polycrystalline materials [5, 24] and as small single crystals [25, 26]. In Ref. 22 we have reported the successful synthesis of non-superconducting and oxygen stoichiometric $RuSr_2GdCu_2O_8$ in 1% of oxygen at 935°C (sample (a) in Fig. 3), i.e., at temperatures considerably lower than usual reported synthesis at 1060°C in flowing oxygen. Superconductivity was then gradually induced in this material (with $T_c$ up to approx. 40 K) by prolonged oxygen annealing (approx. 140h) at 1060°C followed by slow (1°C/min) cooling to room temperature (sample (b) in Fig.3) [22]. When the same superconducting sample of $RuSr_2GdCu_2O_8$ is re-annealed at 1060°C, but then quenched to room temperature, superconductivity vanishes (sample (c) in Fig. 3). The temperature of the magnetic transition for superconducting $RuSr_2GdCu_2O_8$ was always observed to be lower than for its non-superconducting counterpart (130-132K vs. 136K - see Fig. 3 for the temperature dependencies of *ac* susceptibility measured for samples (a), (b) and (c)). The increase of $T_c$ correlates with the decrease of $T_N$. Measurements of the thermopower indicated an increase of hole doping for $RuSr_2GdCu_2O_8$ in converting from non-superconductor to superconductor (open and closed circles in Fig. 1) [18]. Studies of heterovalent substitutions in $Ru_{1-x}M_xSr_2GdCu_2O_8$ (M=$Nb^{5+}$, $Sn^{4+}$) reveal that a decrease of $T_N$ by approx. 5 K can be accomplished by substitution of 2.5% of $Nb^{5+}$ or 1% of $Sn^{4+}$ in the Ru sublattice [21]. Thus, not only hole doping but also a magnetic dilution effect should be considered as a cause of the differences presented here. The thermogravimetry analysis, probing the bulk of the material, did not indicate meaningful changes in oxygen concentration between superconducting and non-superconducting samples (oxygen content $8 \pm 0.02$ per unit formula, see also Ref. 15). However, one should note that high pressure oxygen annealing, which is required for synthesis of $x \neq 0$ phases, readily stabilizes superconductivity with comparatively high $T_{c1}=40$ K (see open circles in Fig. 3, also [1, 27]). The increase of superconducting $T_c$ with Cu→Ru substitution discovered for $x \neq 0$ samples suggests similar local substitution effects should be carefully investigated for possibility of stabilization of the superconducting phase in nominally stoichiometric $RuSr_2GdCu_2O_8$. Note that when considering the chemically modified local microstructure of the material, the non-uniform variations in oxygen content could certainly influence its superconducting properties, in resemblance to the effect found for $Ru_{0.6}Sr_2GdCu_{2.4}O_{8-\delta}$, which becomes non-superconducting after annealing at 800°C in Ar.

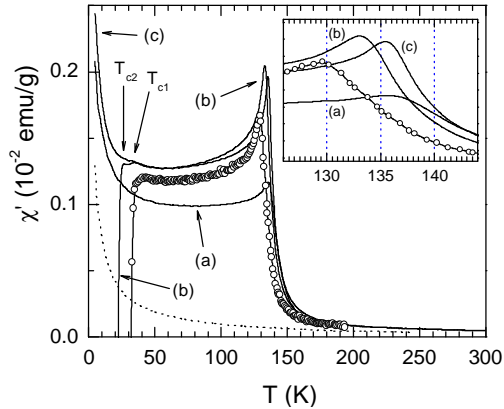

Fig. 3. The ac susceptibility (f=200 Hz, $H_{ac}$=1 Oe) of $RuSr_2GdCu_2O_8$: (a)→(b)→(c) represent the sequence of annealings which convert the same material from NSC to SC and then again to NSC. (a): 1%$O_2$ at 935°C; (b) 140 h $O_2$ at 1060°C slow cooling; (c) after quenching from $O_2$ at 1060°C. Open circles: SC sample after annealing of the NSC precursor (1%$O_2$ at 935°C) in 600bar of oxygen at 1100°C. Dotted line: paramagnetism of $Gd^{3+}Ba_2Cu_3O_{6.2}$. See Ref. 18 for differences between $T_{c1}$, and $T_{c2}$ (onset for the bulk screening) in the $x \neq 0$ series.

Detailed knowledge of the compositional and structural uniformity influenced by differing routes of material processing remains crucial to further understanding the complex behavior of $RuSr_2GdCu_2O_8$.

Contrary to its well-known Ba-based 123-type analogue, superconducting $YSr_2Cu_3O_7$ ($T_c$=60K) was synthesized only at very high pressure (7GPa) of oxygen [28]. The $Ru_{1-x}Sr_2GdCu_{2+x}O_{8-\delta}$ compounds provide an interesting example of stabilization of the Sr-based 123-type structure by substitution of Ru into the Cu(1)-chain sites. Partial substitutions of Cu by different metals M=Fe, Ti, Al, Co, Ga, Pb, Nb, Ta, Mo were found to stabilize the $RESr_2Cu_{3-x}M_xO_{7-\delta}$ structure at ambient pressure conditions [29, 30]. This has been explained as a result of the improvement of lattice matching between the Cu(M)-O chain layer and Sr-O structural layer. Xiong *at al*. [31] reported high-pressure studies of $RESr_2Cu_{2.7}Mo_{0.3}O_{7-\delta}$ (27 K<$T_c$<37 K) and a linear increase of $T_c$ with $dT_c/dP$=7 K/GPa for small RE=Ho, Er, Tm and Yb, and with approx. 4 K/GPa for the larger Tb, Gd, Eu and Sm. The effect can be explained by increase of charge transfer to the $CuO_2$ planes. Additionally, the reduction of the hole carrier concentration, usually being induced by substitution of Cu(1)-chain site, can be at least partially restored under pressure. The underdoped character of Ru$_{1-x}$Sr$_2$GdCu$_{2+x}$O$_{8-\delta}$ samples (0<x<0.5, Fig. 1) suggests that the application of an external pressure should also lead to an increase of T$_c$.

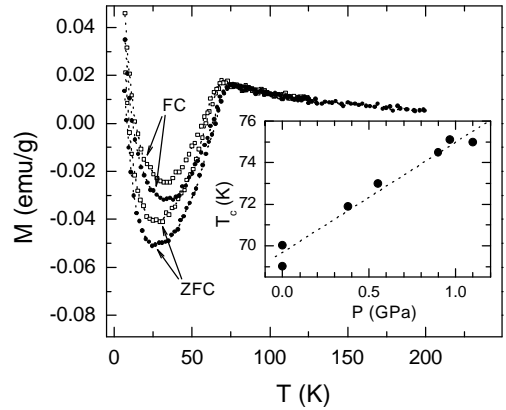

Fig. 4. Temperature dependencies of the dc magnetization for $Ru_{0.6}Sr_2GdCu_{2.4}O_{8-\delta}$ measured at ambient pressure (open symbols) and at 0.97 GPa (full symbols), $H_{dc}$=100 Oe. Inset presents $T_c$ dependence on the external pressure.

Figure 4 presents the zero field cooled (ZFC) and field cooled (FC) magnetization for Ru$_{0.6}$Sr$_2$GdCu$_{2.4}$O$_{8-\delta}$ measured at ambient pressure and at 0.97 GPa. The low temperature reentrance of magnetization reflects the paramagnetic contribution of the Gd ions (for more detailed discussion see Ref. 14, and also Ref. 18 for comparison with properties of isostructural Ru$_{0.6}$Sr$_2$EuCu$_{2.4}$O$_{8-\delta}$, where the paramagnetic contribution of the rare–earth becomes negligible). The inset to Fig. 4 shows T$_c$ versus external pressure for the x=0.4 sample. For all four measured compositions (x=0.1, 0.2, 0.3, 0.4) the T$_c$ increases linearly with pressure at a rate of approx. 5K/GPa. The similar value of dT$_c$/dP observed for both M=Ru and Mo doped GdSr$_2$Cu$_{2+x}$M$_{1-x}$O$_y$ [31] indicate a similarity in the charge transfer mechanism occurring for both compounds. For RuSr$_2$GdCu$_2$O$_8$ Lorenz *at al.* report a positive linear shift of the superconducting T$_c$ at a rate of 1K/GPa and interpret this small rate as resulting from the adverse influence of the weak-ferromagnetic state, for which the ordering temperature increases at a much higher rate of dT$_N$/dP=6K/GPa [32]. Considerably higher rate of dT$_c$/dP≅5K/GPa observed for the series of x≠0 compounds, for which no magnetic order of the Ru sublattice is observed [14], seems to support this interpretation. Our results for the parent x=0 compound show dT$_c$/dP≅1K/GPa and dT$_N$/dP≅6K/GPa, also in agreement with Ref. 32.

The similarity between the crystal structures of RuSr$_2$RE$_{2-y}$Ce$_y$Cu$_2$O$_{10-\delta}$ (1222-type) and Ru$_{1-x}$Sr$_2$GdCu$_{2+x}$O$_{8-\delta}$ (1212-type) compounds (for 1222-type the fluorite type RE(Ce)-O layer replaces a single Gd layer in the 1212-type structure) suggests that high pressure oxygen annealing should also be attempted for stabilization of Cu→Ru substitution in the 1222-type ruthenocuprate. Figs. 5a and 5b present the temperature dependencies of the *ac* susceptibility for two series of Ru$_{1-x}$Sr$_2$Eu$_{2-y}$Ce$_y$Cu$_{2+x}$O$_{10-\delta}$ (y=0.5, 1, and 0 ≤ x ≤ 0.6) after annealing in 600 bar of oxygen at 1100°C. Although minor secondary phases (predominantly CeO$_2$) were still present in several samples after one high pressure annealing, both series reveal a remarkable feature: the Cu→Ru doping gradually diminishes the magnetic response and lowers the temperature of the magnetic transition. For 1222-type composition with the nominal y=0.5 and x=0.4 we report a superconducting transition at T$_{c1}$=48 K (T$_{c2}$=42 K) with no magnetic component detectable in *ac* susceptibility above this temperature. Note that for the 1222-type compound the hole doping can be varied by changing the RE$^{3+}$/Ce$^{4+}$ ratio [2], as well by changes of δ [4, 33]. The T$_c$ dependence on δ for Cu→Ru doped 1222-type phases remains to be investigated. Similar to the x≠0 1212-type compounds, the oxygen content for samples presented in Fig. 5 reflects identical synthesis conditions and thus could vary with x. The differences in magnetic behavior between the y=0.5 and y=1 series are consistent with reduced hole doping induced by increased Ce$^{4+}$→Eu$^{3+}$ substitution. Our recent zero-field μSR experiments for y=0.5, x≥0 samples reveal that the magnetic response (seen here in the *ac* susceptibility above T$_c$, Fig. 5) loses bulk character for higher x. Also, the volume fraction of the sample which responds magnetically diminishes with increasing temperature. Such response can indicate presence of magnetic clusters in the material. Clustering of the Ru atoms should be taken into account for detailed studies of Cu→Ru substituted ruthenocuprates. Detailed analysis of the μSR results is in progress and will be reported separately.

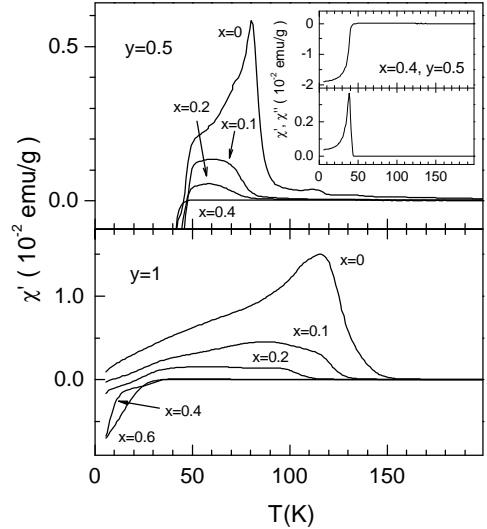

Fig. 5. Temperature dependencies of the ac susceptibility (f=200 Hz, H$_{ac}$=1 Oe) for two series (y=0.5 and y=1) of Ru$_{1-x}$Sr$_2$Eu$_{2-y}$Ce$_y$Cu$_{2+x}$O$_{10-\delta}$ (0≤x≤0.6) as synthesized in 600 bar of oxygen at 1100°C.

In general, it seems that when discussing the properties of ruthenocuprates one should be aware of possible microscale phase separation effects, which may bring new perspectives to the interpretation of the complex properties of these materials.

**Acknowledgments**

Work at NIU was supported by the National Science Foundation (DMR-0105398), and by the State of Illinois under HECA.

S.M.M. acknowledges support by NSF (CHE-9871246 and CHE-9522232). Use of the Advanced Photon Source was supported by the U.S. Department of Energy, Basic Energy Sciences, Office of Science, under Contract No. W-31-109-Eng-38.

I.F. acknowledges financial support by the Israel Science Foundation (2000).

A.W., R.P., and I. M. F acknowledge the Polish State Committee KBN contract No. 5 P03B 12421.

P.W.K., A.S., R.K., I.M.S., B.D. acknowledge work with Dr. D. Herlach and Dr. A. Amato of PSI for the μSR experiments. Authors would like to thank Dr. G. Crabtree and Dr. J. Jorgensen for fruitful discussions.